\documentclass[a4paper, 12pt]{article}
\usepackage{arxiv}
\pdfoutput = 1

\usepackage{tikz}
\usepackage{amsmath}
\usepackage{hyperref}       
\usepackage{url}
\hypersetup{
    colorlinks=true,
    linkcolor=blue,
    citecolor=blue,
    filecolor=blue,      
    urlcolor=blue,
}
\usepackage{booktabs}       
\usepackage{amsfonts}       
\usepackage{nicefrac}       
\usepackage{lipsum}
\usepackage{graphicx}
\usepackage{color,pxfonts,fix-cm}
\usepackage{latexsym}
\usepackage[mathletters]{ucs}
\usepackage[T1]{fontenc}
\usepackage[utf8x]{inputenc}
\usepackage{pict2e}
\usepackage{wasysym}
\usepackage{caption}
\usepackage{authblk}
\usepackage[english]{babel}
\usepackage{ulem}
\usepackage{cancel}
\everymath{\textstyle}
\DeclareMathOperator{\sech}{sech}
\DeclareMathOperator{\arcsinh}{arcsinh}
\DeclareMathOperator{\arccoth}{arccoth}

\usepackage{mathtools}
\begin{document}
\renewcommand\Affilfont{\fontsize{11}{10}\itshape}

\title{\textbf{Relativistic Transformation of Spherical Co-ordinates ($t,r,\theta,\phi$) }}
\author{
 Sarbajit Mazumdar$^1$\thanks{ sarbajit.mazumdar@niser.ac.in} and Krishna Kant Parida $^1$\thanks{ krishna.parida@niser.ac.in}\\
 $^1$School of Physical Sciences\\
 $^1$ National Institute of Science Education and Research, Bhubaneswar, HBNI, P.O. Jatni, Khurda-752050, Odisha, India.\\

}

\maketitle

\begin{abstract}
With the advent of relativistic mechanics, the Lorentz transformation replaced the Galilean transformation based on classical Newtonian mechanics among inertial frames at high uniform velocities, but both transformations are based on Cartesian coordinate system, hence position of particles obtaining linear velocities in space can be obtained. In case where frames are  rotating with constant angular velocity, use of Galilean rotational transformation (GRT) is replaced by Franklin transformation, proposed by Philip Franklin in 1922. The modified transformation introduced the concept of rotational motion of points in a rigid body. Both the transformations are based on cylindrical coordinate system. Here we moved a step further for making a relativistic transformation using spherical coordinate system for understanding the behaviour of rotating frames along any axis in the space passing through the center of mass of a symmetrical object (Sphere). We finally came to an understanding about how Special Theory of relativity is found to be applicable in rotational motion using different co-ordinate system.
\end{abstract}

\section{Introduction}
Rotational motion came into little significance in relativistic mechanics, due to prevalence of the concept of motion of a particle with velocity less than the speed of light $c$. The “velocity” discussed in relativistic mechanics is of rectilinear type, i.e., the rate of change of the magnitude of the component related to the co-ordinates of the Cartesian co-ordinate system, with respect to time. For example, $v_x$ is the rate of change of magnitude of component say $x$, in $\hat{x}$ direction with respect to time t. Similarly, the rate of change of magnitudes $y$ and $z$ in $\hat{y}$ and $\hat{z}$ direction respectively with respect to time $t$, is
given by $v_y$ and $v_z$ respectively. One of the important aspects of the Cartesian co-ordinate system is that all the components are of same dimension, i.e. the length. But with the development of the concept of rotational motion, more forms of co-ordinate systems were introduced, whose components need not have same dimensions, which are called Curvilinear co-ordinates. Cylindrical and spherical co-ordinate systems are the examples. In Cylindrical co-ordinates, the radial $\hat{s}$ and $\hat{z}$ components have same dimension of length but the angular $\hat{\phi}$ component is not of that dimension. Similarly for spherical co-ordinates, $\hat{\theta}$ and $\hat{\phi}$ components have same dimensions for angle, but not of length which is of the $\hat{r}$ component.

Hence, much of the transformations for relative frames of references, based on the rectilinear co-ordinates were able to successfully describe both classical and special relativity by Galilean and Lorentz transformation respectively. But here come the question, what about the particle revolving around a fixed point with a fixed ”velocity”. In this case we cannot define velocity with respect to the usual definitions used in rectilinear co-ordinate system,
but we can easily do that with a curvilinear co-ordinate system. Hence, the Galilean transformation for cartesian co-ordinate system was extended to both spherical and cylindrical co-ordinate systems under the name of Galilean Rotational Transformation(GRT), which is discussed in the next section. With time, in agreement with the concept of special relativity, the
GRT was replaced by Franklin Transformation for cylindrical coordinates and we tried to formulate how to make another transformation with agreement to special relativity for spherical co-ordinates.

\section{{Galilean Transformation of Rotating Frames}}
To explain the Galilean transformation, we consider two different frames one is at rest and other one is rotating with constant angular velocity $\Omega$ measured by the inertial observer. Using spherical coordinates, we denote the space-time coordinates of the non-rotating frame by \textbf{$(t,r,\theta,\phi)$} and the rotating one by \textbf{$(t',r',\theta',\phi')$}. We can take two types of rotation one is along  $\phi$ direction and  another one is along the $\theta$-direction.

\subsection{\textbf{For rotation of frame along \texorpdfstring{$\phi$}{Lg} direction :}}
\label{lab1}
 
Let's take two observers present in two frames $S$ and $S'$ having a common origin ($O$). The frame $S'$ is rotating around $S$ frame with a uniform angular velocity $\Omega_{\phi}$. Let's take an arbitrary point $P$ at a distance $r$ from the origin, in the space, and this point is observed from both of the frames.  

\vspace{2mm}  
From the Figure \ref{fig:galilean1}, since the frame $S'$ is rotating along $\phi$ direction, both frames $S$ and $S'$ share a common $Z$-axis. Let the coordinates of the point $P$ be {$(t,r,\theta,\phi)$} for $S$ frame and {$(t',r',\theta',\phi')$} for $S'$ frame. Hence, the coordinates of point $P$ in both frames are related by:

\begin{figure}[ht]
\begin{center}

\begin{tikzpicture}[scale = 0.64, transform shape ]
\draw [green, ultra thick,->] (0,0)--(0,8.1) ;
\draw [red,ultra thick,->]  (0,0)--(-4,-4);
\draw[blue,ultra thick,->] (0,0)--(6,0);

\draw [black,thick, <- , domain = 420:220] plot ({1*cos(\x)},{6.5+0.2*sin(\x)});
\draw [green, ultra thick,->] (0,0)--(0,8.1) ;
\draw [red,dashed,ultra thick,->]  (0,0)--(4,-4);
\draw [blue,dashed,ultra thick,->]  (0,0)--(5.3,2.2);
\draw [black,ultra thick,->] (0,0)--(2.5,4.2);
\draw [black,dashed,ultra thick,->](2.5,4.2)--(2.5,-1.2);
\draw[black](0,0)--(2.5,-1.2);
\draw [black,thick, <- , domain =335:225] plot ({2.1*cos(\x)},{2.1*sin(\x)});
\draw [black,thick, <- , domain = 315:225] plot ({1.2*cos(\x)},{1.2*sin(\x)});
\draw [black,thick, <-> , domain = 334:316] plot ({2.6*cos(\x)},{2.6*sin(\x)});
\draw [black,thick, <-> , domain = 91:59] plot ({1.9*cos(\x)},{1.9*sin(\x)});

\node at (6.2,0.0) {Y};
\node at (0,8.4) {Z};
\node at (-4.2,-4.2) {X};
\node at (1.4,6.7) {\Large {$\Omega_{\phi}$}};
\node at (5.6,2.3) {$Y'$};
\node at (4.4,-4.4) {$X'$};
\node at (-0.2,0.3) {O}; 
\node at (0,-0.9) {\Large{$\Omega t$}};
\node at (0,-2.4) {\Large{$\phi$}};
\node at (2.4,-1.8) {\Large{$\phi'$}};
\node at (3.4,4.5) {\Large{P$(r'=r)$}};
\node at (2.8,-1.1) {\Large{$R$}};
\node at (0.3,1.1) {\Large{$\theta$}};

\end{tikzpicture}
\end{center}
\caption{Rotation of co-ordinate axis along $\phi$ direction}
\label{fig:galilean1}
\end{figure}
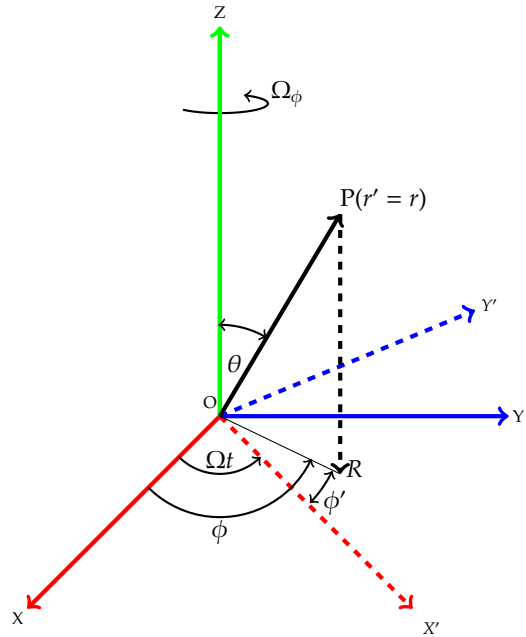

\begin{subequations}
\begin{align}
 t' &= t\\
r'&=r\\
\phi'&= \phi-\Omega_{\phi}t \\
\theta' &= \theta 
\end{align}
\end{subequations}

\vspace{1.5 mm}
Now differentiating both sides of the equations (1b), (1c) and (1d) with respect to $t$, 

\begin{subequations}
\begin{align}
\frac{dr'}{dt}&=\frac{dr}{dt}\\
\frac{d\phi'}{dt}&= \frac{d\phi}{dt}-\Omega_{\phi} \\
\frac{d\theta'}{dt} &= \frac{d\theta}{dt} 
\end{align}
\end{subequations}

 Since there is no movement along $r$ direction and no rotation along $\theta$ direction.

Hence,

\begin{subequations}
\begin{align}
&\frac{dr'}{dt}=\frac{dr}{dt}=0\\
&\frac{d\theta'}{dt} = \frac{d\theta}{dt}=0
\end{align}
\end{subequations}

Now, let $\omega_{\phi}$ and $\omega_{\phi}'$  are the angular velocities observed in frame $S$ and $S'$ respectively.

Thus,

\begin{subequations}
\begin{align}
&~~\qquad\frac{d\phi'}{dt}=\frac{d\phi}{dt}-\Omega_{\phi}\\
&\implies\boxed{{\omega_{\phi}'} = {\omega_{\phi}}-\Omega_{\phi}}
\end{align}
\end{subequations}

 The above equation is previously discussed in \cite{franklin}.

\subsection{{\textbf{For rotation of frame along \texorpdfstring{$\theta$}{Lg} direction :}}}
\label{lab2}
Using conditions similar to $\phi$ rotation, let $OP$ makes an angle $\theta$ with $Z$-axis of $S$ frame and now the $S'$ frame starts rotating along the $\theta$ direction with uniform angular velocity ${\Omega_{\theta}}$. 

\vspace{2mm}  
From the Figure \ref{fig:galilean}, since the frame $S'$ is rotating along $\theta$ direction. Let the coordinates of the point $P$ be {$(t,r,\theta,\phi)$} for $S$ frame and {$(t',r',\theta',\phi')$} for $S'$ frame. Hence, the coordinates of point $P$ in both frames are related by:

\begin{figure}[ht]
\begin{center}

\begin{tikzpicture}[scale = 0.62, transform shape ]

\draw [green, ultra thick,->] (0,0)--(0,7) ;
\draw [red,ultra thick,->]  (0,0)--(-4,-4);
\draw[blue,ultra thick,->] (0,0)--(7,0);
\draw [green,dashed, ultra thick,->] (0,0)--(3.4,5.7);
\draw [black,thick, <-> , domain = 91:50] plot ({1.9*cos(\x)},{1.9*sin(\x)});
\draw [black,thick, -> , domain = 91:59] plot ({3.3*cos(\x)},{3.3*sin(\x)});
\draw [black,thick, <-> , domain = 60:50] plot ({4*cos(\x)},{4*sin(\x)});
\draw [red, dashed,ultra thick,->]  (0,0)--(-2.8,-4.3);
\draw[blue,dashed,ultra thick,->] (0,0)--(5.8,-2);
\draw [black,ultra thick,->] (0,0)--(3.6,4.2);
\draw [black,dashed,ultra thick,->](3.5,4.2)--(3.4,-2.2);
\draw [black,thick](0,0)--(3.4,-2.2);
\draw [black,dashed,ultra thick,->](3.5,4.2)--(2.1,-3.2);
\draw [black,thick](0,0)--(2.1,-3.2);
\draw [black,thick,<->, domain =328:225] plot ({1.9*cos(\x)},{1.9*sin(\x)});
\draw [black,thick,<-> ,domain =303:238] plot ({3.3*cos(\x)},{3.3*sin(\x)});
\draw [black,thick, -> , domain =96:55] plot ({1.9+5.5*cos(\x)},{1.2+5.7*sin(\x)});

\node at (7.3,0.0) {Y};
\node at (0,8) {Z};
\node at (-4.2,-4.2) {X};

\node at (0.8,2.3) {\Large{$\theta$}};
\node at (0.9,3.6) {\Large {$\Omega_{\theta}t$}};
\node at (-2.1,-4.6) {$X'$};
\node at (5.8,-2.5) {$Y'$};
\node at (3.6,6) {$Z'$};
\node at (4.5,4.4) {\Large P{$(r'=r)$}};
\node at (2.45,3.6) {\Large{$\theta'$}};
\node at (-0.2,0.3) {O}; 
\node at (3.4,-2.6) {\Large{$R$}};
\node at (2.1,-3.6) {\Large{$R'$}};
\node at (0,-1.3) {\Large{$\phi$}};
\node at (0,-3.9) {\Large{$\phi$}};
\node at (3.6,7){\Large {$\Omega_{\theta}$}};

\end{tikzpicture}
\end{center}
\caption{Rotation of co-ordinate axis along $\theta$ direction}
\label{fig:galilean}
\end{figure}
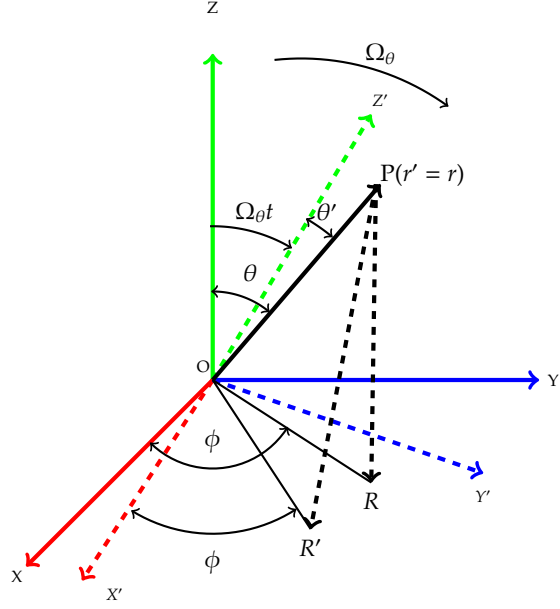

    \begin{subequations}
\begin{align}
    t' &= t\\
r'&=r\\
\theta'&= \theta-\Omega_{\theta}t \\
\phi' &= \phi 
\end{align}
\end{subequations}

Now equation (5d) may look non trivial. Since $\phi$ is the angle between $X$-axis and the horizontal projection of $OP$, this angle is not changing while rotating the $S'$ frame along $\theta$. So we can conclude $\phi'=\phi$.   

\vspace{2mm}
Similarly, let $\omega_{\theta}$ and $\omega_{\theta}'$  are the angular velocities observed in frame $S$ and $S'$ respectively.

Thus,

    \begin{subequations}
\begin{align}&~~\qquad\frac{d\theta'}{dt}=\frac{d\theta}{dt}-\Omega_{\theta}\\
&\implies\boxed{{\omega_{\theta}'} = {\omega_{\theta}}-\Omega_{\theta}}
\end{align}
\end{subequations}

This equation is also previously discussed in \cite{franklin}
\vspace{2mm}

\section{\textbf{Relativistic Transformation of Rotating Frames}}

 For any relativistic transformation, like that of Lorentz or that of Franklin  \cite{df}, following conditions are required to be fulfilled:
\vspace{2mm}

 1. If the point $P$ is revolving with respect to moving frame along either $\phi$ or $\theta$ direction, the magnitude of the tangential velocity must be independent of time, and is same for all the points at same distance from the axis of rotation.
 \vspace{2mm}
 
 2. The equation of transformation must have the possibility to reduce into the equations for Galilean transformation, i.e. the magnitude of tangential velocity will be much less than the speed of light $c$.

\subsection{\textbf{Transformation for boost along \texorpdfstring{$\phi$}{Lg} direction : }}
 To describe the rotation, we can compare the frame $S'$ rotating with a constant angular velocity $\Omega_{\phi}$, having an axis along $Z$-direction. The point $P$ is fixed with respect to $S$ frame.
The equation of transformations must be similar to any linear Lorentz boost, with a difference that the linear distance is replaced by an arc length of the disc ${l_{\phi}=r_s \phi}$, where $r_s$ is perpendicular distance of the point $P$ from the axis of rotation($Z$ axis) and we have $r_{s}=r\sin(\theta)$.

Considering the above conditions, following transformation laws can be obtained:

    \begin{subequations}
\begin{align}&t' =k(r_s)\bigg{( t-\frac{v(r_s)r_s\phi}{c^2}\bigg)}\\
&r'=r\\
&\theta'= \theta\\
&r_s'\phi' = k(r_s)(r_s\phi-v(r_s)t)
\end{align}
\end{subequations}

\textbf{Matrix form :} We can express these equations by using a single matrix equation.

\begin{center}
          
$\begin{bmatrix}
ct' \\
r' \\
r'\theta'\\
r_s'\phi'
\end{bmatrix}= \begin{bmatrix}
k(r_s) & 0 & 0 & -\beta k(r_s) \\
0 & 1 & 0 & 0 \\
0 & 0 & 1 & 0\\
-\beta k(r_s) & 0 & 0 & k(r_s)
\end{bmatrix}\begin{bmatrix}
ct \\
r \\
r\theta\\
r_s\phi
\end{bmatrix}$
\end{center} 

\vspace{2mm}

where, $k(r_s)=\frac{1}{\sqrt{1-\frac{(v(r_s))^{2}}{c^{2}}}}$ is the Lorentz-type factor, $v(r_s)$ is the tangential velocity of point $P$ observed in frame $S'$ and $\beta = \frac{v(r_s)}{c}$.

\subsection{\textbf{{Transformation for boost along \texorpdfstring{$\theta$}{Lg} direction : }}}
\label{lab3}
To describe this rotation, we can compare the frame $S'$ rotating with a constant angular velocity $\Omega_{\theta}$, having an axis along $(\pmb{{r}} \times \pmb{\hat{z}})$-direction through the origin($O$), which is co-planar to $XY$ plane. Similar to previous case here is also the point $P$ is fixed.
Here, the linear distance is replaced by an arc length of the disc ${l_{\theta}}=r\theta$ and as the axis of rotation  is passing through the origin we can say $r_s=r$ .
\vspace{1mm}

 Similarly, another set of transformation laws are obtained:

    \begin{subequations}
\begin{align}&t' =k(r_s)\bigg{( t-\frac{v(r_s)r\theta}{c^2}\bigg)}\\
&r'=r\\
&r'\theta' = k(r_s)(r\theta-v(r_s)t)\\
&\phi'= \phi
\end{align}
\end{subequations}

\textbf{Matrix form :} Similarly here also, we can represent these equations by using a single matrix equation.
\vspace{2mm}

\begin{center}
    
$\begin{bmatrix}
ct' \\
r' \\
r'\theta'\\
r_s'\phi'
\end{bmatrix}= \begin{bmatrix}
k(r_s) &0 & -\beta k(r_s)& 0 \\
0 & 1& 0 & 0 \\
-\beta k(r_s) & 0 & k(r_s) & 0\\
0 & 0 & 0 & 1
\end{bmatrix}\begin{bmatrix}
ct \\
r \\
r\theta\\
r_s\phi
\end{bmatrix}$
\end{center}

\section{\textbf{Expression For Tangential Velocity  for Spherical Co-ordinates}}

\subsection{\textbf{{Expression for \texorpdfstring{$r$}{Lg} in \texorpdfstring{$\phi$}{Lg} rotation : }}}

\begin{figure}[ht]
\begin{center}

\begin{tikzpicture}[scale = 0.7, transform shape ]

\draw [green, ultra thick,->] (0,0)--(0,8.1) ;
\draw [red,ultra thick,->]  (0,0)--(-4,-4);
\draw[blue,ultra thick,->] (0,0)--(6,0);
\draw [black ,thick, ->] (0,0)--(2,2.236);
\draw [black ,dashed, <->] (2,2.236)--(0,5);
\draw (2,2.236)--(0,2.27);
\draw (0,2.26) ellipse (2 cm and 0.2 cm);
\draw (0,0) circle  (3cm);
\draw [ black, dashed, <-> ] (-1,0)--(-1,5);
\draw [black,thick, <- , domain = 420:220] plot ({1*cos(\x)},{6.5+0.2*sin(\x)});
\draw [black,thick, <-> , domain = 91:46] plot ({1*cos(\x)},{1*sin(\x)});

\node at (6.5,0.0) {Y};
\node at (1.4,4.0) {\Large{$r_{\phi}$}};
\node at (1.4,6.7) {\Large {$\Omega_{\phi}$}};
\node at (1.5,1.2) {\Large{$r$}};
\node at (1,2.5) {\Large{{$r_{s}$}}};
\node at (-1.5,1.8) {\Large{$z_{\phi}$}};
\node at (0.5,1.3) {\Large {$\theta$}};
\node at (0,8.4) {Z};
\node at (-4.2,-4.2) {X};
\node at (-0.6,5) {A};
\node at (-0.6,2.27) {B};
\node at (0,-0.5) {O};
\node at (2.3,2.29) {P};

\end{tikzpicture}
\end{center}
\caption{ Rotation along $\phi$ direction }
\label{fig: rotation}

\end{figure}
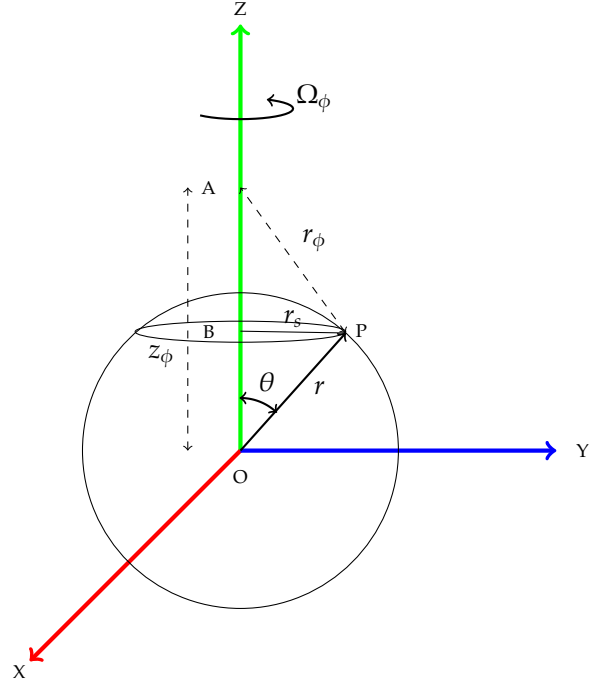

According to Figure \ref{fig: rotation},  let's consider a sphere of radius $r$ centered at origin ($O$) passing through point $P$. Now suppose an observer is located at $A$ on $Z$ axis at a distance $z_{\phi}$ from origin. Let $r_{\phi}$ be the distance form the observer to point $P$. Now consider a cross-section of the sphere of radius $r_s$, parallel to $XY$ plane, passing through point $P$.

From the Figure \ref{fig: rotation}, using Cosine rule in ${\triangle {AOP}}$ we get,
$$r_{\phi}^{2}= r^{2}+z_{\phi}^{2}-2rz_{\phi}\cos{\theta}$$

By using, $r=r_s\csc(\theta)$ 
we obtain a Quadratic equation,$$r_{s}^{2}\csc^{2}(\theta)-2z_{\phi}r_{s}\cot(\theta)+(z_{\phi}^{2}-r_{\phi}^{2})=0 $$ 

By using Quadratic equation formula we get, $$ r_s = \sin(\theta)\bigg[z_{\phi}\cos(\theta)+\sqrt{r_{\phi}^{2}-z_{\phi}^{2}\sin^{2}(\theta)}\bigg]$$

Hence, the relation between  $z_{\phi}$, $r_{\phi}$ and $r$ is given by$$\boxed{r=\bigg[z_{\phi}\cos(\theta)+\sqrt{r_{\phi}^{2}-z_{\phi}^{2}\sin^{2}(\theta)}\bigg]}$$

So we can find the value of $r$ for any arbitrary position of the observer on $Z$ axis for any  general point  on the surface of the sphere.

\subsection{\textbf{{Expression for \texorpdfstring{$v(r_s)$}{Lg} in \texorpdfstring{$\phi$}{Lg} rotation : }}}

let's consider a point $P$ be fixed on the sphere of radius $r$, and both the point and the sphere (attached to $S'$ frame) are rotating along $\phi$ direction with a constant angular velocity of $\Omega_{\phi}$. In this case, unlike the conditions mentioned in section \ref{lab1} , here  we are observing the point $P$ attaining tangential velocity $v(r_s)$ in $S$ frame. Since in both conditions, frames are moving with an constant angular velocity $\Omega_{\phi}$, the observed tangential velocity $v(r_s)$ in both the cases will be \textbf{equal}.

 Now to obtain the expression of $v(r_s)$ of the point $P$ in spherical co-ordinates, we have to consider the following Figure \ref{rotation}.

 \begin{figure}[ht]
\begin{center}

\begin{tikzpicture}[scale = 0.68, transform shape ]

\draw [green, ultra thick,->] (0,0)--(0,10.3);
\draw (0,3.5) ellipse (1.9 cm and 0.5 cm);
\draw(0,5.2) ellipse (2.8 cm and 0.5cm);
\draw [green, ultra thick,] (0,3.5)--(0,4.1);
\draw [green, ultra thick,] (0,5.2)--(0,5.8);
\draw [black,thick,dashed,<->] (1.95,3.59)--(2.8,5.2);
\draw [black,thick,->] (0,0)--(1.9,3.5);
\draw [black,thick,->] (2,3.5)--(0.8,4.4);
\draw [black,thick,->](2.9,5.2)--(0.8,6.75);
\draw [black,ultra thick](0,9)--(0,9.1);
\draw [black,thick, <-> , domain = 91:60] plot ({1.5*cos(\x)},{1.5*sin(\x)});
\draw [black,thick,<->] (0,3.5)--(1.9,3.5);
\draw [black,thick,<->] (0,5.2)--(2.8,5.2);
\draw (0,0) circle  (4.13cm);
\draw (0,0) circle  (5.96cm);
\draw [red,ultra thick,->]  (0,0)--(-6,-6);
\draw[blue,ultra thick,->] (0,0)--(10,0);

\node at (2, 6.8) {{$v(r_s+\Delta{r_s})$}};
\node at (1.3,4.5){{$v(r_s)$}};
\node at (0,-0.4) {O};
\node at (0,10.7) {Z};
\node at (2.9,4.2) {$\Delta{r}$};
\node at (-0.7,9.1) {A};
\node at (1.5, 2.1) {$r$}; 
\node at (0.5,1.9) {$\theta$};
\node at ((-0.5,3.5) {$B$};
\node at ((-0.5,5.2) {$B'$};
\node at ((2.5,3.45) {$P$};
\node at ((3.4,5.2) {$P'$};
\node at (0.9,3.3) {$r_s$};
\node at (0.9,5.45) {$r_{s}+\Delta{r_s}$};
\node at (10.3,0.0) {Y};
\node at (-6.2,-6.2) {X};

\end{tikzpicture}
\end{center}
\caption{Direction of $v(r)$ and $v(r+\Delta r)$ in $\phi$ rotation}
\label{rotation}

\end{figure}
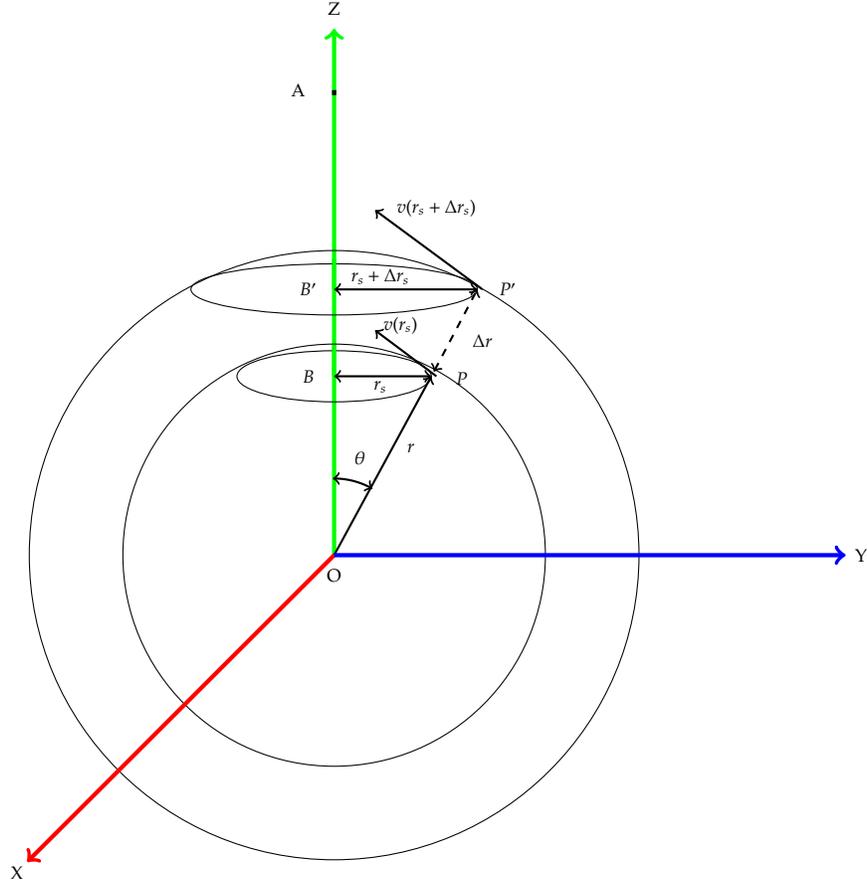

Now if we displace the point $P$ to $P'$ radially by very small amount i.e. $\Delta r$. The point $P'$ will be on an another sphere of radius $r+\Delta r$, and its tangential velocity will be $v(r_s+\Delta r_s)$. Since points $P$ and $P'$ are collinear both the velocities $v(r_s)$ and  $v(r_s+\Delta r_s)$ are parallel to each other.
\vspace{2mm}

From the figure $\triangle{B'OP'}$ and $\triangle{BOP}$ are similar we can say 
$$\frac{r}{r_s}=\frac{r+\Delta r}{r_s+\Delta r_s}$$
$$\implies \frac{\Delta r_s}{\Delta r}=\frac{r_s}{r}= \sin(\theta)$$
\vspace{2mm}

As mentioned in section \ref{lab1} the equation of transformation of $\phi$ co-ordinate is given by,

    \begin{subequations}
    \begin{align}&r_s'\phi' = k(r)(r_s\phi-v(r_s)t)
\end{align}
\end{subequations}

Taking differential on both sides,

    \begin{subequations}
\begin{align}
& d(r_s'\phi') = k(r)\bigg(d(r_s\phi) - v(r_s)dt\bigg)
\end{align}
\end{subequations}

Now, taking the equation for transforming time($t$) co-ordinate

    \begin{subequations}
    \begin{align}&t' =k(r)\bigg{( t-\frac{v(r_s)r_s\phi}{c^2}\bigg)}
\end{align}
\end{subequations}

Taking differential on both sides,

    \begin{subequations}
    \begin{align}&dt' =k(r)\bigg{( dt-\frac{v(r_s)d(r_s\phi)}{c^2}\bigg)}
\end{align}
\end{subequations}

Now, by dividing equations (10a) \& (12a) we get,

\begin{align*}&~~~~~~\frac{d(r_s'\phi')}{dt'} = \frac{\cancel{k(r)}\bigg(d(r_s\phi) - v(r_s)dt\bigg)}{\cancel{k(r)}\bigg{(dt-\frac{v(r_s)d(r_s\phi)}{c^2}\bigg)}}
\end{align*}

\begin{subequations}
\begin{align}&\implies \frac{d(r_s'\phi')}{dt'} = \frac{\bigg(\frac{d(r_s\phi)}{dt} - v(r_s)\bigg)}{\bigg{(1-\frac{v(r_s)\frac{d(r_s\phi)}{dt}}{c^2}\bigg)}} 
\end{align}
\end{subequations}

Now, $\frac{d(r_s\phi)}{dt}$ is the tangential velocity of point $P'$ observed from frame $S$ and $\frac{d(r_s'\phi')}{dt'}$ is the tangential velocity of point $P'$ observed from  point $P$ in $S'$ frame.

\vspace{1mm}

Hence, $\frac{d(r_s\phi)}{dt} = v(r_s+\Delta r_s)$ and $\frac{d(r_s'\phi')}{dt'} = {\Omega_{\phi}(r_s+\Delta r_s-r_s)} = {\Omega_{\phi}(\Delta r_s)}$
\vspace{2mm}

Hence, the above mentioned equation (13a) will be changed into
\begin{align*}&\Omega_{\phi}(\Delta r_s) = \frac{v(r_s+\Delta r_s)-v(r_s)}{1-\frac{v(r_s+\Delta r_s)(v(r_s))}{c^{2}}}\\
&\\ 
&\implies \Omega_{\phi}=\frac{ \frac{v(r_s+\Delta r_s)-v(r_s)}{\Delta r_s}}{1-\frac{v(r_s+\Delta r_s)(v(r_s))}{c^{2}}}
\end{align*}

Now we know that $\Delta r_s \rightarrow 0$ So,

\begin{align*}&\implies \Omega_{\phi}=\lim_{\Delta r_s \rightarrow 0}\frac{ \frac{v(r_s+\Delta r_s)-v(r_s)}{\Delta r_s}}{1-\frac{v(r_s+\Delta r_s)(v(r_s))}{c^{2}}}\\[15pt]
&\implies \Omega_{\phi}=\frac{\frac{dv(r_s)}{dr_s}}{1-\frac{v^{2}(r_s)}{c^{2}}} \\[15pt] 
&\implies \frac{\Omega_{\phi}{dr_s}}{c^2}=\frac{dv(r_s)}{c^2-v^{2}(r_s)} 
\end{align*}
Integrating both sides,
\begin{align*}&\implies \frac{\Omega_{\phi}}{c^2}{\int_{0}^{r_s} dr_s}=\int_{0}^{v(r_s)}\frac{dv(r_s)}{c^2-v^{2}(r_s)}\\[15pt]
&\implies \frac{\Omega_{\phi}{r_s}}{c^2}=\ln{\bigg[\frac{c+v(r_s)}{c-v(r_s)}\bigg]}\times{\frac{1}{2c}}\\[15pt]
&\implies \bigg[\frac{c+v(r_s)}{c-v(r_s)}\bigg]=e^{\frac{2\Omega_{\phi}{r_s}}{c}}\\[15pt]
&\implies  v({r_s})=c\Bigg[ \frac{e^{\frac{2\Omega_{\phi}r_s}{c}}-1}{e^{\frac{2\Omega_{\phi}r_s}{c}}+1}\Bigg]
\end{align*}

since, $r_s=r\sin(\theta)$, we get the final expression

$$
 \boxed{v({r\sin(\theta}))=c\Bigg[ \frac{e^{\frac{2\Omega_{\phi}r\sin(\theta)}{c}}-1}{e^{\frac{2\Omega_{\phi}r\sin(\theta)}{c}}+1}\Bigg]}
$$
\\
\\

\subsection{\textbf{{Expression for \texorpdfstring{$r$}{Lg} \& \texorpdfstring{$v(r_s)$}{Lg} in \texorpdfstring{$\theta$}{Lg} rotation : }}}
Similar to section \ref{lab3}, let's consider a point $P$ be fixed on the periphery of a incompressible disc of radius $r$, centered at $O$ and both the point and the disc (attached to $S'$ frame) are rotating along $\theta$ direction with a constant angular velocity of $\Omega_{\theta}$. The axis of rotation is $A'B'$, along $(\pmb{{r}} \times \pmb{\hat{z}})$-direction through the origin($O$),  as mentioned in section \ref{lab2}. In this case, unlike the conditions mentioned in section \ref{lab2} , as the frame $S'$ and the point $P$ are moving together, we observe that point $P$ is attaining tangential velocity $v(r_s)$ in $S$ frame. Since in both conditions, frames are moving with an constant angular velocity $\Omega_{\theta}$, the observed tangential velocity $v(r_s)$ in both the cases will be \textbf{equal}.

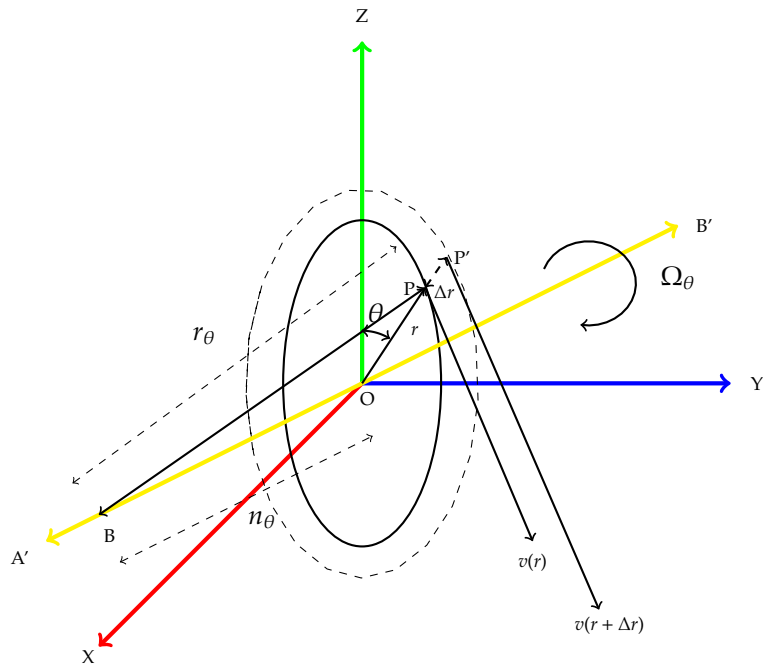
\begin{figure}[ht]
\begin{center}

\begin{tikzpicture}[scale = 0.7, transform shape ]

\draw [green, ultra thick,->] (0,0)--(0,6.5) ;
\draw [red,ultra thick,->]  (0,0)--(-5,-5);
\draw[blue,ultra thick,->] (0,0)--(7,0);
\draw[black,thick] (0,0) ellipse (1.5 cm and 3.1 cm);
\draw [black,dashed, domain = 560:150] plot ({2.21*cos(\x)},{3.7*sin(\x)});

\draw [yellow,ultra thick, <-] (-6,-3)--(0,0);
\draw [yellow,ultra thick, ->] (0,0)--(6,3);
\draw [black, thick, ->] (0,0)--(1.2,1.82);
\draw [black,thick, <->] (-5,-2.5)--(1.2,1.82);
\draw [black,thick, <-> , domain = 91:58] plot ({1*cos(\x)},{1*sin(\x)});
\draw [black,thick, -> , domain = 160:-100] plot ({4.3+0.9*cos(\x)},{1.9+0.8*sin(\x)});
\draw [black,dashed, <->] (-5.5,-1.9)--(0.65,2.6);
\draw [black,dashed, <->] (-4.6,-3.4)--(0.2,-1.0);
\draw [black, thick,dashed,<->] (1.2,1.82)--(1.6,2.4);
\draw [black,thick,-> ] (1.2,1.82)--(3.24,-3);
\draw [black,thick,-> ] (1.6,2.4)--(4.5,-4.3);

\node at (7.5,0.0) {Y};
\node at (0,7) {Z};
\node at (-5.2,-5.2) {X};
\node at (-6.5,-3.3) {A'};
\node at (6.5,3) {B'};
\node at (-3,0.9) {\Large{$r_{\theta}$}};
\node at (-1.9,-2.6) {\Large{$n_{\theta}$}};
\node at (0.1,-0.3) {O};
\node at (0.9,1.8) {P};
\node at (1.9,2.4) {P'};
\node at (1,1) {$r$};
\node at (3.24,-3.4) {$v(r)$};
\node at (4.7, -4.6) {$v(r+\Delta r)$};
\node at (1.56,1.75) {$\Delta r$};
\node at (0.26,1.35) {\Large{$\theta$}};
\node at (-4.8,-2.9) {B};
\node at (6,2) {\Large{$\Omega_{\theta}$}};

\end{tikzpicture}
\end{center}
\caption{Direction of $v(r)$ and $v(r+\Delta r)$ in $\theta$ rotation }
\label{fig: rotation1}

\end{figure}

We can generalise this for any position of the observer on $A'B'$ axis.
According to Figure \ref{fig: rotation1} if the observer is at $B$ point and the distance of point $P$ to $B$ is $r_{\theta}$ and let $n_{\theta}$ be the distance of point B from origin. Then, $r=\sqrt{r_{\theta}^{2}-n_{\theta}^2}$.
\vspace{0.6mm}

In this case the expression of $v(r_s)$ will be completely similar to that of $\phi$ rotation except here $r_s=r$ as the $r_s$ is distance of point $P$ from the axis $A'B'$.  $$
 \boxed{v(r_s)=v(r)=c\Bigg[ \frac{e^{\frac{2\Omega_{\theta}r}{c}}-1}{e^{\frac{2\Omega_{\theta}r}{c}}+1}\Bigg]}
$$

\section{\textbf{Final Transformation Laws For Rotating Frames}}

We have the general equation of $v(r_s)$ as $$\displaystyle{v({r_s})=c\Bigg[ \frac{e^{\frac{2\Omega r_s}{c}}-1}{e^{\frac{2\Omega r_s}{c}}+1}\Bigg]}$$

we know that $\displaystyle{{\tanh{x} = \frac{e^{2x}-1}{e^{2x}+1}}}$ ~~~~\cite{Hyper}
 
Hence, $$\displaystyle{v({r_s})=c~ \tanh\Bigg(\frac{\Omega r_s}{c}}\Bigg)$$

The value of $k(r_s)$ is given by, $$~~~~~~~{ k(r_s)=\frac{1}{\sqrt{1-\frac{(v(r_s))^{2}}{c^{2}}}}}$$
$$~~~~~~~\implies k(r_s)=\frac{1}{\sqrt{1-\frac{(c \tanh({\frac{\Omega r_s}{c}}))^{2}}{c^2}}}$$
$$~~~~~~~\implies k(r_s)= \frac{1}{\sqrt{1-(\tanh{\frac{\Omega r_s}{c}})^{2}}}$$ 
$$\implies k(r_s) =\cosh{\bigg(\frac{\Omega r_s}{c}\bigg)}$$
Now, $$\beta k(r_s) = \tanh\Bigg(\frac{\Omega r_s}{c}\Bigg)\times \cosh{\bigg(\frac{\Omega r_s}{c}\bigg)}= \sinh{\bigg(\frac{\Omega r_s}{c}\bigg)}$$ 

So, the final matrix equation for $\phi$ rotation will be :

\begin{center}
    
$\begin{bmatrix}
ct' \\
r' \\
r'\theta'\\
r_s'\phi'
\end{bmatrix}= \begin{bmatrix}
\cosh \left(\frac{\Omega r_{s}}{c}\right) &0 &0& -\sinh \left(\frac{\Omega r_{s}}{c}\right)\\
0 & 1& 0 & 0 \\
0 & 0 & 1 & 0\\
-\sinh \left(\frac{\Omega r_{s}}{c}\right) & 0 & 0 & \cosh \left(\frac{\Omega r_{s}}{c}\right)

\end{bmatrix}\begin{bmatrix}
ct \\
r \\
r\theta\\
r_s\phi
\end{bmatrix}$
\end{center}

The final matrix equation of $\theta$ rotation will be :

\begin{center}
    
$\begin{bmatrix}
ct' \\
r' \\
r'\theta'\\
r_s'\phi'
\end{bmatrix}= \begin{bmatrix}
\cosh \left(\frac{\Omega r_{s}}{c}\right) &0 & -\sinh \left(\frac{\Omega r_{s}}{c}\right)& 0 \\
0 & 1& 0 & 0 \\
-\sinh \left(\frac{\Omega r_{s}}{c}\right) & 0 & \cosh \left(\frac{\Omega r_{s}}{c}\right) & 0\\
0 & 0 & 0 & 1
\end{bmatrix}\begin{bmatrix}
ct \\
r \\
r\theta\\
r_s\phi
\end{bmatrix}$
\end{center}

\section{\textbf{Expression For Centripetal Acceleration}}

Let $a{(r_s)}$ be the centripetal acceleration of point $P$ in both cases( $\theta$ and $\phi$ rotation), then expression for  $a_{(r_s)}$ will be :

\begin{align*}
&~~~~~~~~~~~ a{(r_s)}=\frac{dv(r_s)}{dt}= v(r_s)\frac{d(v(r_s))}{dr_s}\\
&\implies a{(r_s)}= c ~\tanh{\bigg(\frac{\Omega r_s}{c}\bigg)}\frac{d\bigg(c \tanh{\bigg(\frac{\Omega r_s}{c}\bigg)}\bigg)}{dr_s}\\
&\implies \boxed{a{(r_s)} = \Omega c~ \tanh{\bigg(\frac{\Omega r_s}{c}\bigg)} \sech^{2}{\bigg(\frac{\Omega r_s}{c}\bigg)}}
\end{align*}
\pagebreak
\section{\textbf{Reduction of Relativistic Transformation to Galilean  Approximation}}

Since, We have the equation,  $$\displaystyle{v({r_s})=c~ \tanh\Bigg(\frac{\Omega r_s}{c}}\Bigg)$$ 

We are using Taylor approximation of the functions to reduce it.

Let,$$f(x)= \tanh{(x)}$$
So, Taylor Approximation of the function is 
$$ f(x) \approx P(x) = x-\frac{x^3}{3}+\frac{2x^5}{15}- \cdots $$  
So whenever, $$\frac{\Omega r_s}{c} << 1$$
we can say 
\begin{align*}&~~~~~~~~~~~ v(r_s)= c~ \tanh\Bigg(\frac{\Omega r_s}{c}\Bigg)\\
&\implies v(r_s)= c~\bigg[\frac{\Omega r_s}{c}-\frac{\Omega^{3} {r_s}^3}{3c^3}+ \cdots \bigg]\\
&\implies v(r_s)= \cancel{c} \times \frac{\Omega r_s}{\cancel{c}}\\
&\implies \boxed{ v(r_s)= \Omega r_s}
\end{align*}

Again for the equation of acceleration $$a{(r_s)} = \Omega c~ \tanh{\bigg(\frac{\Omega r_s}{c}\bigg)} \sech^{2}{\bigg(\frac{\Omega r_s}{c}\bigg)}$$
 So Let $$f(x)= \tanh{(x)}\sech^{2}{(x)}$$
 Hence the Taylor expansion of the function follows
 $$ f(x) \approx P(x) = x-\frac{4x^3}{3}+\frac{17x^5}{15}- \cdots $$  
So whenever, $$\frac{\Omega r_s}{c} << 1$$
we can say \begin{align*}&~~~~~~~~~ a(r_s)= \Omega c~ \tanh\Bigg(\frac{\Omega r_s}{c}\Bigg)\sech^{2}{\bigg(\frac{\Omega r_s}{c}\bigg)}\\
&\implies a(r_s)= \Omega c~\bigg[\frac{\Omega r_s}{c}-\frac{4\Omega^{3} {r_s}^3}{3c^3}+ \cdots \bigg]\\
&\implies a(r_s)= \Omega \cancel{c} \times \frac{\Omega r_s}{\cancel{c}}\\
&\implies \boxed{ a(r_s)= \Omega^{2} r_s}
\end{align*}
So,  for both of the final expression of $v(r_s)$ and $a(r_s)$ have been reduced to the classical form. 

\section{\textbf{Graphical Nature}}

So we have the expression,  $$\displaystyle{v({r_s})=c~ \tanh\Bigg(\frac{\Omega r_s}{c}}\Bigg)$$ 
Now the following graph represent the variation of $v(r_s)$ with $r_s$.

\begin{figure}[ht]
    \centering
    \includegraphics[scale=0.6]{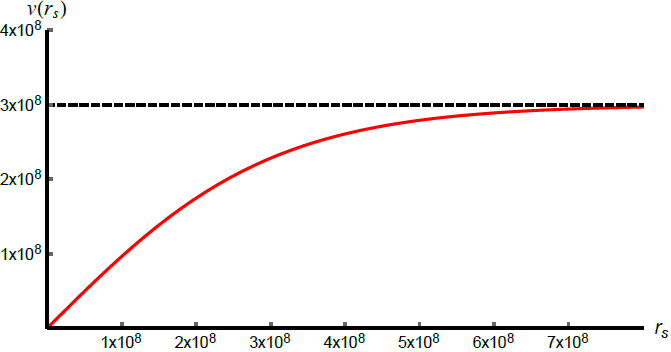}
    \caption{Variation of $v(r_s)$ vs. $r_s$}
    \label{fig:my_v(r_S)}
\end{figure}

From the graph (Figure \ref{fig:my_v(r_S)}) we can see that there is a horizontal asymptote at $c = 3 \times 10^8 $m/s , we can conclude that $v(r_s)$ will never go beyond $c$. Here we have taken $\Omega=1$ rad/s, but for any values of $\Omega$, $v(r_s)$ will never cross $c$. 

Now for Acceleration, we have
$$a{(r_s)}=\frac{dv(r_s)}{dt}= v(r_s)\frac{d(v(r_s))}{dr_s}$$

For the equation for acceleration, we can say that the magnitude of centripetal acceleration is directly proportional to the product of $v(r_s)$ and slope of the graph between $v(r_s)$ and $r_s$. Now initially, since the equation of $v(r_s)$ behaves similar to that for Galilean transformation, the slope of the graph remains constant and $a(r_s)$ increases with respect to the value of $v(r_s)$. However, after a particular point, the slope of the graph decreases and hence after a certain point, it dominates over the increasing value of $v(r_s)$ resulting in decrease in the value of $a(r_s)$. Finally, as $r_s \rightarrow \infty$, $v(r_s) \rightarrow c$ but $\frac{dv(r_s)}{dr_s} \rightarrow 0$ and therefore, $a(r_s) \rightarrow 0$.
\begin{figure}[ht]
    \centering
    \includegraphics[scale = 0.6]{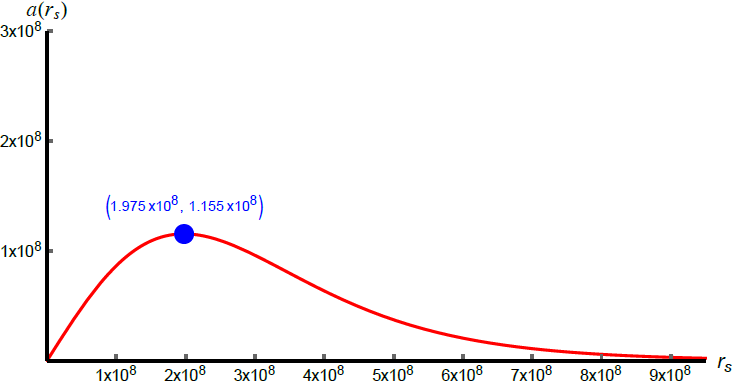}
    \caption{Variation of $a(r_s)$ vs. $r_s$}
    \label{fig:my_label10}
\end{figure}
\pagebreak
We can also calculate at which value of $r_s$, $a(r_s)$ attains maximum.

\begin{align*}&~~~~~~~~\frac{d[a(r_s)]}{dr_s}=0\\
&\implies 2 \tanh^{2}{\bigg(\frac{\Omega r_s}{c}\bigg)}- \sech^{2}{\bigg(\frac{\Omega r_s}{c}\bigg)} =  0\\
&\implies \sinh^{2}{\bigg(\frac{\Omega r_s}{c}\bigg)}= \frac{1}{2}\\
&\implies \frac{\Omega r_s}{c} = \arcsinh{\bigg(\frac{1}{\sqrt{2}}\bigg)}\\
&\implies \frac{\Omega r_s}{c} = \ln{\bigg(\frac{1+\sqrt{3}}{\sqrt{2}}\bigg)}\bigg[ \text{ As we know that } \arcsinh{(x)}=\ln{\bigg(x+\sqrt{x^2 +1}\bigg)}\bigg]\\
&\implies\boxed{ r_s = \frac{c}{\Omega}~\ln{\bigg(\frac{1+\sqrt{3}}{\sqrt{2}}\bigg)}}
\end{align*}

Here we have taken $\Omega = 1 $ rad/s so we have $r_s=3 \times 10^{8}\times\ln{\bigg(\frac{1+\sqrt{3}}{\sqrt{2}}\bigg)}= 1.975 \times 10^{8} $ m

Now as mentioned before after a particular point the slope of the graph of $v(r_s)$ vs $r_s$ decreases, we have found that the third order derivative i.e. $\frac{d^{3}(v(r_s)}{dr_s^{3}}=0$ at $r_s = 1.975 \times 10^{8}$ m, if $\Omega = 1$ rad/s. That means after $r_s = 1.975 \times 10^{8}$ m as $r_s$ increases $v(r_s) \rightarrow c$ and $a(r_s) \rightarrow 0$.

Now for value of $r_s$, the corresponding value of $v(r_s)$ at $\Omega = 1 $ rad/s is given by
\begin{align*}&~~~~~~~~~~~v(r_s) = c \times \tanh{\bigg(\frac{r_s}{c}\bigg)} \text{m/s}\\
&\implies v(r_s) = c \times \tanh{\bigg[\ln{\bigg(\frac{1+\sqrt{3}}{\sqrt{2}}\bigg)}\bigg]}~ \text{m/s}\\
&\implies v(r_s) = c \times \frac{1} {\coth{\bigg[\ln{\bigg(\frac{1+\sqrt{3}}{\sqrt{2}}\bigg)}\bigg]}} \text{ m/s}\\
&\implies v(r_s) = c \times \frac{1} {\coth{\bigg[\ln{\bigg(\sqrt{\frac{(1+\sqrt{3})^{2}}{2}}\bigg)}\bigg]}}~\text{m/s}\\
&\implies v(r_s) = c \times \frac{1} {\coth{\bigg[\frac{1}{2}\ln{\bigg(\frac{\sqrt{3}+1}{\sqrt{3}-1}\bigg)}\bigg]}}~\text{m/s}\\
&\implies v(r_s) = \frac{c}{\sqrt3} ~\text{m/s} = 1.732 \times 10^{8}~\text{m/s} ~~ \text{\bigg(As we know that,}~\arccoth{(x)}= \frac{1}{2} \ln{\bigg(\frac{x+1}{x-1}\bigg)}~\forall~x>1   \bigg)
\end{align*}
Hence, we can say that when the magnitude of $v(r_s)$ reaches value of $1.732 \times 10^{8}$ m/s or around $57.7\%$ of the speed of light $c$, the magnitude of tangential acceleration $a(r_s)$,  attains peak value, beyond which, it decreases to zero.
\pagebreak

\section{Lorentz Invariance of proven relativistic transformation }
The theory of Lorentz Invariance says that, the laws of physics are invariant under
a transformation between two co-ordinate frames moving at constant constant velocity w.r.t. to each other i.e. the transformation we formulated here should be invariant at both frame of references. Whenever we talk about the transformation of spherical co-ordinates from \textbf{$(ct,r,r\theta,r_s\phi)$} to \textbf{$(ct',r',r\theta',r_s\phi')$}, the value four dimensional length, let's say that the space-time interval, denoted by $S^2$ should be invariant at both frame of reference.\\
Therefore we have to prove that,
 $$ S^2 = (ct')^2-(r')^2-(r'\theta')^2-(r_s'\phi')^2 = (ct)^2-(r)^2-(r\theta)^2-(r_s\phi)^2 $$
 
 Now we take the transformation of $\phi$ Co-ordinate,   
\begin{center}
    
$\begin{bmatrix}
ct' \\
r' \\
r'\theta'\\
r_s'\phi'
\end{bmatrix}= \begin{bmatrix}
\cosh \left(\frac{\Omega r_{s}}{c}\right) &0 &0& -\sinh \left(\frac{\Omega r_{s}}{c}\right)\\
0 & 1& 0 & 0 \\
0 & 0 & 1 & 0\\
-\sinh \left(\frac{\Omega r_{s}}{c}\right) & 0 & 0 & \cosh \left(\frac{\Omega r_{s}}{c}\right)

\end{bmatrix}\begin{bmatrix}
ct \\
r \\
r\theta\\
r_s\phi
\end{bmatrix}$
\end{center} 

Now following the transformation laws we have,
\begin{align*}& S^2 = (ct')^2-(r')^2-(r'\theta')^2-(r_s'\phi')^2 \\
&~~~ = \left[\cosh \left(\frac{\Omega r_{s}}{c}\right)ct-\sinh \left(\frac{\Omega r_{s}}{c}\right)r_s\phi \right]^{2}-(r)^2-(r\theta)^2- \left[ -\sinh \left(\frac{\Omega r_{s}}{c}\right)ct + \cosh \left(\frac{\Omega r_{s}}{c}\right)r_s\phi \right]^{2}\\
\\
&~~~ = \left[\cosh^{2} \left(\frac{\Omega r_{s}}{c}\right)c^2 t^2 + \sinh^{2} \left(\frac{\Omega r_{s}}{c}\right)r_s^{2}\phi^{2}- 2\cosh \left(\frac{\Omega r_{s}}{c}\right)\sinh \left(\frac{\Omega r_{s}}{c}\right)ctr_s\phi\right] -(r)^2-(r\theta)^2\\
&~~~~~~~~~~~~~~~~~~~~~~~~~~~~~~~~~~~~~~~~~~-\left[ \sinh^{2} \left(\frac{\Omega r_{s}}{c}\right)c^2 t^2 +\cosh^{2} \left(\frac{\Omega r_{s}}{c}\right)r_s^{2}\phi^{2}- 2\cosh \left(\frac{\Omega r_{s}}{c}\right)\sinh \left(\frac{\Omega r_{s}}{c}\right)ctr_s\phi \right]\\
\\
&~~~  = c^2 t^2\left[\cosh^{2} \left(\frac{\Omega r_{s}}{c}\right)- \sinh^{2} \left(\frac{\Omega r_{s}}{c}\right)\right] - (r)^2-(r\theta)^2 - r_s^{2}\phi^{2}\left[\cosh^{2} \left(\frac{\Omega r_{s}}{c}\right)- \sinh^{2} \left(\frac{\Omega r_{s}}{c}\right)\right]\\
&~~~ = (ct)^2-(r)^2-(r\theta)^2-(r_s\phi)^2 ~~\text{\bigg(As we know that,}~\cosh^{2}(x)- \sinh^{2}(x) = 1 \bigg)
    \end{align*}
So, we can see that the transformation of $\phi$ Co-ordinate is Lorentz Invariant. Similarly, we can show that the transformation for $\theta$ Co-ordinate will also be Lorentz Invariant.

\section{Validity of Special Theory of Relativity in Curvilinear co-ordinates}

Applicability of Special Theory of Relativity (STR) is not centralized upon accelerated motion to be considered, it is based upon how you define acceleration in a coordinate system. Since Cartesian coordinate system is rectilinear, acceleration is simply defined as rate of change of velocity along any component of x-axis, y-axis and z-axis respectively. In other words, it is also defined as the double derivative of displacement along any component with respect to time. STR is easily applicable since all is there up to when there is no velocity varying with time.

Now it is true that circular motion is accelerated and involves centrifugal force, but the real cause of acceleration is due to change in direction rather than change in magnitude of velocities. In rectilinear or Cartesian coordinate system, the circular motion seems accelerated because the components of velocities in any direction are varying with time. But in curvilinear coordinate system under our consideration, all components of velocities are constant and do not vary with time. Hence STR to that extent in curvilinear system is applicable, but not in Cartesian coordinate system. As referring to the Famous quote by Feynman \cite{feynman}, “There is no relativity of rotation”.

It is not possible to keep a free moving body or particle to have all the velocity components constant with time in a curvilinear coordinate system without a radial force acting on it, due to tendency of free moving body to undergo curvilinear motion. Hence force and acceleration values will be available as calculated, but those will be just the cause of motion, rather than affecting the motion itself. If we think the  applicability of General Theory of Relativity (GTR), it will be still difficult to understand how centripetal as well as centrifugal force acting on the particle under observation (present on the sphere) will affect the space-time as well as the path of the object with time since we didn't specify the cause and type of force responsible for circular motion of the body and irrespective of its type, the trajectory of the particle remains same.

In case of light, one direct conclusion is that due to its rectilinear nature of propagation, it’s not possible to completely define the path in terms of curvilinear coordinate system. Any particle undergoing circular motion with the absolute speed of light will experience no acceleration along radial direction, which means the circular motion is inconsistent in that speed, as the net force along the radial direction is zero. Hence in conclusion, no photon can be confined around a single point.  The above conclusion seems to be refuted that Light can never be confined around a single point, as we know that Fibre optics can change the direction of light, making a curvilinear trajectory.

\begin{figure}[ht]
    \centering
    \includegraphics[scale = 0.6]{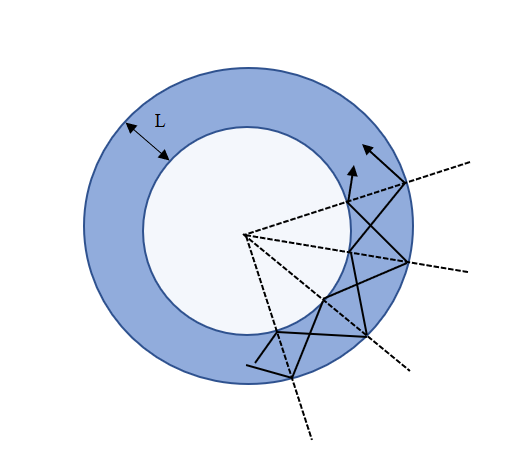}
    \caption{Light confined in a Fiber Optic Cable}
    \label{fig:my_label}
\end{figure}

From the above Figure \ref{fig:my_label}, it can be easily seen that light can be made to move around a point, i.e. the center of the fibre optic cable. In the cable, we observe that photons are undergoing successive total internal reflections. Hence the effective tangential velocity of the photons is less than the absolute speed of light. Also the photons are still undergoing rectilinear propagation, but along a curved path. Now suppose we consider the width of the fibre optic cable to be very small, comparable to the dimension of the photon as a particle. With that, it can be argued that the effective tangential velocity of the photon will be made equal to the speed of light.

But here is a catch. Suppose let $\rho_0$ be the radius of the photon. Then we can have the width $L$ of the fibre optic cable. Let light consist of two photons only, and they are just moving along the fiber optic cable. Now let the system is analogous to  "Two" particles in a box\cite{wave} of width $L=2\rho_{0}$. Hence the Wave function of each particle is given by, $\Psi_{n}(x)= \sqrt{\dfrac{1}{\rho_{0}}} \sin\left(\dfrac{n\pi x}{2\rho_{0}}\right)$. So Expectation value of position of each photon $\left<x\right>$ is given by\\
\begin{align*}&~~~~~~~~~~~\left<x\right> = \int_{0}^{2\rho_{0}} \Psi_{n}^{*}(x)x\Psi_{n}(x)dx\\
&\implies \left<x\right> = \int_{0}^{2\rho_{0}}\dfrac{1}{\rho_{0}}x\sin^{2}\left(\dfrac{n\pi x}{2\rho_{0}}\right)dx\\
&\implies \left<x\right> = \rho_{0}
    \end{align*}
    
Since each photon is expected to be found at $x=\rho_{0}$ , we can say that both of the photons are expected to be found at the same position at any given interval of time, which is completely absurd according to the dimension of the fibre optic cable. Hence, our assumption for taking the width of the optic cable to the dimension of the size of photon is invalid and hence the effective tangential velocity of the photon can never be equal to the speed of light $c$.

\section{\textbf{Conclusion}}
We have discussed in our paper a proposed relativistic rotational transformation for spherical coordinates from existing Franklin transformation based on cylindrical coordinates. In this, we related coordinates of an inertial static frame to the frame rotating around a common axis with constant angular velocity $\Omega_\phi$ and $\Omega_\theta$ along direction of azimuth ($\phi$) and zenith ($\theta$) angle respectively. We also discussed how the actual tangential velocity and angular velocity are related with respect to this relativistic transformation. Although it looks similar to Franklin transformation, but while introducing the concept of this proposed rotational transformation, we tried to use certain standard physical quantities like distance of the point from the rotational axis, denoted by $r_s$ and its significance with changing direction of rotation. Based on it, a new form of rotational kinematics can be obtained, which always reduces to Galilean approximations when the magnitude of tangential velocity is around its classical limits.
\vspace{2mm}

We also discussed the behaviour of both tangential velocity and centripetal acceleration with increasing value of distance of the point under observation, revolving around an axis with fixed angular velocity $\Omega$. Another thing we can say is that in actual, since magnitude of tangential velocity $v(r_s)$ is dependent upon a function of  $\Omega$ and $r_s$ , similar conclusions can also be drawn for varying magnitude of $\Omega$ and a fixed distance from the rotating axis. We can also conclude that the formulations are consistent with its physical significance, concluding that any massive particle reaching speed of light can never be completely confined in a rotating system, since the centripetal acceleration responsible for its confinement always reduces to zero. Now we also considered a possible system to refute the argument about confinement of light around a single point using example of fibre optic cable. But also proved how this system is consistent with the given argument. Finally, we provided an explanation about how STR is valid for curvilinear co-ordinates, even with the presence of forces and accelerated motion.

\section{Acknowledgement}
We would like to thank {Dr.~Najmul Haque} of the School of Physical Sciences, NISER for many useful discussions and insightful comments and also theoretical guidance. S.M. would like to acknowledge the support of INSPIRE(SHE) Programme of Department of Science and Technology, Govt. of India. K.P. would also like to acknowledge the support of KVPY Programme, funded by Department of Science and Technology, Govt. of India and implemented by IISc, Bangalore.  We would also like to thank the Academic administration of NISER.

\end{document}